\documentclass[prb,showpacs,preprint,aps]{revtex4}
\usepackage{graphicx}
\usepackage{amsmath}
\usepackage{amssymb}
\usepackage{epsfig}
\begin{document}

\title{Plasmon Enhanced Faraday Rotation in Thin Films}

\author{Zhyrair Gevorkian$^{1,2,*}$ and Vladimir Gasparian$^{3}$}
\affiliation{
$^{1}$ Yerevan Physics Institute,0036 Yerevan, Armenia. \\
$^{2}$ Institute of Radiophysics and Electronics,Ashtarak-2,0203,Armenia.\\
$^{3}$ California State University, Bakersfield\\
$^{*}$ gevork@yerphi.am}




\begin{abstract}
We have analyzed analytically the Faraday rotation of an electromagnetic wave for magnetoactive thin metallic film with a nanostructured surface profile. Periodical as well as random surface profiles were considered. The plasmon contribution to the Faraday angle was studied. For periodical grating case, we have shown that the maximum of rotation angle is achieved when surface plasmon wave number coincides with one of the wave numbers of the inverse lattice. Enhancement of the Faraday angle at plasmonic band edges is predicted. In the case of random surface profile, it is shown that the diffusion of surface magnetoplasmons gives a dominant contribution to Faraday rotation. Comparison with the experiments is carried out.
\end{abstract}


\maketitle
 \section{Introduction}
In recent years
many experimental papers on enhanced Faraday rotation in the systems with nanoscale inhomogeneities have appeared \cite{uchida09}-\cite{nature13}. This increased research interest is largely motivated by the fact that the Faraday effects
is widely used in optical isolators, phase modulators \cite{Hay08, Yu08}, spin dynamics \cite{Liu12} and etc. Experiments found that the origin of the strong enhancement in the classical regime is intimately connected with the different plasmon \cite{Raether88} resonances. The papers \cite{Jain09,Chikan11,Chikan13} deal with the three-dimensional (3D) random systems consisted of
a solution with the embedded in it metallic nanoparticles. In such systems, an enhancement of Faraday rotation at the frequencies close to the nanoparticle surface plasmon resonance frequency is observed in modest magnetic
fields.
Other experimental papers \cite{uchida09,Tkachuk11,nature13} were devoted to
the plasmon induced enhancement of Faraday rotation
when an electromagnetic wave passes through a subwavelength
thin metallic film with nanostructured surface profile.
The surface profile can be in the form of a
periodical grating (as in the experiment \cite{nature13} where inhomogeneity
is created by the periodically placed nanowires
on the surface) as well as random, as in case of randomly
embedded nanoparticles on the surface \cite{uchida09,Tkachuk11}.

The theory, outlined in Ref.\cite{GG13} for Faraday rotation in 3D disordered
media, could correctly predict many of the
peculiar features of the experiments \cite{Jain09,Chikan11, Chikan13}. According to Ref. \cite{GG13} the Faraday rotation angle
in 3D disordered system is inversely proportional to the
photon elastic mean free path, depends on the frequency, and has a minimum at the frequency of nanoparticle local
plasmon resonance, due to a large scattering cross section.

However, most of the experiments on enhanced Faraday rotation are carried out with the subwavelength thin metallic films that can be considered as 2D systems. A consistent theory of Faraday rotation in 2D disordered systems is absent.

 In the present paper, we theoretically consider the Faraday rotation of light passing through a thin metallic film with structured surface profile. Within a common approach we study both periodical and random surface profiles. We show that the plasmon scatterings on the inhomogeneities of surface profile lead to rotation angle enhancement.

 \section{Formulation of the problem.} Let a p-polarized wave impinge on the interface between two media (see Fig.1).
\begin{figure}
\includegraphics[width=8.4cm]{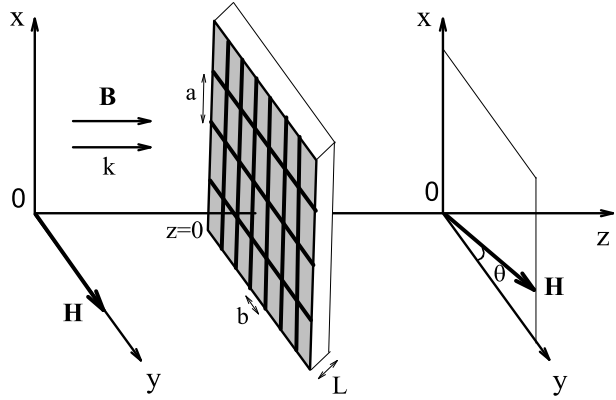}
\caption{Geometry of the problem. Incident wave is polarized
on $0y$.  The external magnetic field is directed on $z$. After passing through a thin film, the incoming beam
is rotated by the Faraday angle $\theta$.}
\label{fig.1}
\end{figure}
The incident wave magnetic field $\vec H$ is directed on $0y$. After passing the magnetooptical medium, it rotates in the plane $xy$. The plane of incidence of wave vector is $xz$. The dielectric permittivity tensor of the system
has the form
 \begin{equation}
 \varepsilon_{ij}(\vec r)=(\varepsilon_0(z)+\varepsilon_s(\vec r))\delta_{ij}-ie_{ijz}g,
 \label{tensor}
 \end{equation}
 where $\varepsilon_0(z)=1$ if $z<0$ and $z>L$ and  $\varepsilon_0(z)=\varepsilon(\omega)$ if $L>z>0$. The term $\varepsilon_0$ describes the smooth surface and $\varepsilon_s(\vec r)=(\varepsilon-1)\delta(z)h(x,y)$ describes the surface roughness. Here $h(\vec \rho)$ is the surface profile ($\vec\rho$ is a two dimensional vector on the plane $xy$) that can be random as well as periodical and $e_{ijk}$ is the antisymmetric tensor. $g$ describes the magneto-optical properties of the medium, and we assume that the external magnetic field is directed on $0z$.  The above geometry is more frequently used in the experiments.
  Faraday rotation angle is determined as
 \begin{equation}
 \tan\theta=\frac{H_x(L)}{H_y(L)},
 \label{faran}
 \end{equation}
where $L$ is the thickness of the film. Assuming that the profile function $h(\vec\rho)$ is smooth and neglecting its derivatives, from Maxwell equations one obtains a Helmholtz equation for the magnetic field
 \begin{equation}
 \triangle H_i(\vec r)+\frac{\omega^2}{c^2}\varepsilon_{ij}(\vec r)H_j(\vec r)=0.
 \label{helmholtz}
 \end{equation}
 Substituting Eq.(\ref{tensor}) into Eq.(\ref{helmholtz}), it is easy see that equations for right-hand and left-hand polarized photons are separated
 \begin{equation}
 \triangle H_{\pm}(\vec r)+k_0^2\varepsilon_{\pm}(\vec r)H_{\pm}(\vec r)=0,
 \label{leftright}
 \end{equation}
 where $k_0=\omega/c$, $H_{\pm}=H_y\pm iH_x$ and $\varepsilon_{\pm}(\vec r)=\varepsilon_0+\varepsilon_s\pm g$. We neglect the difference between left-hand and right-hand polarizations in $\varepsilon_s$ because it is already proportional to a small parameter $h(\rho)$. The Faraday angle is determined through $H_{\pm}$ as
 \begin{equation}
 \tan\theta=-i\frac{H_{+}(L)-H_{-}(L)}{H_{+}(L)+H_{-}(L)}.
 \label{detangle}
 \end{equation}
 Because $H_{x,y}$ are continuous at $z=0$,  the same is correct for $H_{\pm}$. From the Maxwell equations and continuity of $E_{x,y}$ follow the continuity of $(1/ \varepsilon_{\pm})\partial H_{\pm}/\partial z$ at $z=0$.

 \section{Smooth Surface}. When the surface roughness is absent ($\varepsilon_s\equiv 0$) one can solve the Helmholtz equation, Eq.(\ref{helmholtz}), for $\vec H$  with the above mentioned boundary conditions and find
 \begin{equation}
 H_+^0(0)=1+r_{+};\quad H_+^0(L)=t_{+}e^{ik_0L},
 \label{backsol}
 \end{equation}
where the reflection and transmission amplitudes are determined as follows (see, e.g., Ref. \cite{arkady})
\begin{eqnarray}
r_+=\frac{(\varepsilon_+-1)(e^{2ik_0L}-1)}{e^{2ik_0L}(1-\sqrt{\varepsilon_+})^2-(1+\sqrt{\varepsilon_+})^2}\nonumber \\
 t_+=e^{i(k_+-k_0)L}(1+r_+\frac{1-\sqrt{\varepsilon_+}}{1+\sqrt{\varepsilon_+}}).
\label{reftrans}
\end{eqnarray}
Here $k_+=k_0\sqrt{\varepsilon_+}$ and analogous expressions for $r_-$ and $t_-$ can be written substituting $+$ by $-$. Substituting Eqs.(\ref{backsol}) and (\ref{reftrans}) into Eq.(\ref{detangle}) , for the thick films $k_0L\gg 1$ one finds the well known result for the Faraday angle \cite{LL}
 \begin{equation}
 \theta_0=\frac{gk_0L}{2\sqrt{\varepsilon}}.
 \label{angleflat}
 \end{equation}
In the thin film limit $k_0L\ll 1$ similarly, we find
\begin{equation}
\theta_0=\frac{gk_0L}{2},
\label{thin}
\end{equation}
where  we assume that $g\ll|\varepsilon|$. In analogous manner one can find the polarization rotation angle for the reflected wave
\begin{eqnarray}
\tan\theta^R&=&\frac{ig}{1-\varepsilon},\quad k_0L\ll 1\\ \nonumber
  &            &\frac{ig}{(1-\varepsilon)\sqrt{\varepsilon}}, \quad k_0L\gg 1
\label{reflect}
\end{eqnarray}
Note that in both limits the rotation angle does not depend on $L$.

\section { Scattered field}
The solution of the Eq. (\ref{leftright}) consists of two contributions: one is caused by the smooth surface and the second one is caused by the scattering from the inhomogeneities $H_{\pm}(\vec r)=H^0_{\pm}(\vec r)+H^s_{\pm}(\vec r)$, where the background field obeys a homogeneous equation
\begin{equation}
\triangle H^0_{\pm}(\vec r)+k_0^2\varepsilon_{\pm} H^0_{\pm}(\vec r)=0.
\label{back}
\end{equation}
The scattered field $H^s_{\pm}(\vec r)$ is determined through the Green's function
\begin{equation}
H^s_{\pm}(\vec r)=-k_0^2\int d\vec r^{\prime}G_{\pm}(\vec r,\vec r^{\prime})\varepsilon_s(\vec r^{\prime})H^0_{\pm}(\vec r^{\prime}),
 \label{scafield}
 \end{equation}
 where the Green's function obeys the equation
 \begin{equation}
 \triangle G_{\pm}(\vec r,\vec r^{\prime})+k_0^2(\varepsilon_{\pm} +\varepsilon_s(\vec r))G_{\pm}(\vec r,\vec r^{\prime})=\delta(\vec r-\vec r^{\prime}).
 \label{grfunc}
 \end{equation}
Below, we separately consider the case when the surface of metal film has a periodical grating and when the surface profile is random.

\subsection{ Periodical grating}
In this case,  $h(\vec\rho)$ is a two-dimensional periodic function. One can expand the profile function into discrete Fourier series
\begin{equation}
h(\vec \rho)=\sum_{n,m}h_{nm}e^{i\vec K_{nm}\vec \rho}
\label{perprof}
\end{equation}
where $\vec K_{nm}\equiv (2\pi n/a, 2\pi m/b)$ are two dimensional discrete vectors on the inverse lattice; $a,b$ are the profile periods in $x,y$ directions respectively and $n,m=0,\pm 1, \pm 2.... $.  When one of the periods tends to infinity, one recovders a one dimensional periodical profile, considered in the experiment \cite{nature13}. Substituting Eq.(\ref{perprof}) into Eq.(\ref{scafield}), taking its 2D Fourier transforms, and integrating over $z$ using the explicit form of $\varepsilon_s(\vec r)$, one finds
\begin{equation}
H_+^s(\vec r)=-k_0^2(\varepsilon-1)H_+(0)\sum_{nm}h(K_{nm})e^{i\vec K_{nm}\vec\rho}G_+(\vec K_{nm}|Z,0^+),
\label{scafield2}
\end{equation}
where $G_+(\vec p|z,z^{\prime})$ is the two dimensional Fourier transform:
\begin{equation}
G_+(\vec \rho-\vec\rho^{\prime},z,z^{\prime})=\int\frac{d\vec p}{(2\pi)^2}G_+(\vec p|z,z^{\prime})e^{i\vec p(\vec\rho-\vec\rho^{\prime})}.
\label{2dfourier}
\end{equation}
It is worth noticing, that the presence of the $\delta$-function in the expression of $\varepsilon_s$ will lead to the different values of any physical quantity at $ z = 0$, while evaluating the integral over $z$. To avoid the problem with discontinuous physical quantities at $z = 0$ in our further
calculations, we will take their value at $z = 0^+$. One has an analogous expression for the left-hand polarized component. Note that the background field $H_+^0(\vec r)$ which is the solution of homogeneous equation Eq.(\ref{back}) depends only on $z$. In order to obtain the Faraday rotation angle , see Eq.(\ref{detangle}) and Fig.1, we need to evaluate the coherent part of the scattered field, Eq.(\ref{scafield2}), that is the part with wave vector directed on $z$
\begin{equation}
H_+^{sc}(z)=-k_0^2(\varepsilon-1)H_+(0)\sum_{nm}h(K_{nm})G_+(\vec K_{nm}|z,0^+).
\label{cohpart}
\end{equation}
As it is seen from Eq.(\ref{cohpart}), the scattered field includes a Green's function that has a plasmon pole which plays a crucial role in our study of the magneto-optic effects in 2D disordered systems. More precisely, when one of the wave numbers of the inverse lattice $K_{nm}$ coincides with the plasmon wave number then the scattered field resonantly enhances (see below).

In order to analyze the Faraday angle, taking into account the scattered field, let us represent it in the form
\begin{equation}
\tan\theta=-i\frac{H_+^0(L)-H_-^0(L)+H_+^{sc}(L)-H_-^{sc}(L)}{H_+^0(L)+H_-^0(L)+H_+^{sc}(L)+H_-^{sc}(L)}.
\label{faranglesc}
\end{equation}
For the thin films $H_+^0(L)-H_-^0(L)\sim L$. At the resonance $H_+^{sc}(L)-H_-^{sc}(L)$ can be essentially larger than $H_+^0(L)-H_-^0(L)$. At the same time, $H_+^{sc}(L)+H_-^{sc}(L)$ is proportional to the roughness height $h/\lambda$ and is significantly smaller than $H_+^0(L)+H_-^0(L)$. The latter is proportional to unity provided that $L\to 0$. Correspondingly the Faraday angle will resonantly enhance provided that resonance condition is fulfilled. Such an experimental enhancement of Faraday rotation is observed in the recent experiment \cite{nature13}.

It is worth noting that for the reflected wave the above mentioned effect is absent due to the fact that the denominator and numerator of Eq. (\ref{faranglesc}) at the resonance are of the same order, i.e., $H_+^{R0}(0)\sim r_+ \sim L\to 0$.
For the thick films, $k_0L\gg 1$, the resonance effect is possible.

In order to estimate the plasmon contribution to the Faraday rotation angle for thin film, we assume that grating height is small $h\ll \lambda$ and substitute the Green's function in Eq.(\ref{cohpart}) by the bare one $(\varepsilon_s\equiv 0)$.

\subsection{Magnetoplasmon Green's function} For a given $z^{\prime}$, the Green's function has a sense of magnetic field of a point source. Therefore it satisfies the same boundary conditions as magnetic field, namely continuity of $G$ and $(1/\varepsilon_\pm)\partial G_{\pm}/\partial z$. Solving Eq.(\ref{grfunc}) for $\varepsilon_s\equiv 0$ with the above mentioned boundary conditions at $z=0$, one obtains
\begin{equation}
G_+^0(p|z,0^+)=\frac{-ie^{i\sqrt{k_+^2-p^2}Z}}{\sqrt{k_+^2-p^2}+\varepsilon_+\sqrt{k_0^2-p^2}},\quad z>0.
\label{magplasmon}
\end{equation}
It is easy to find the Green's function for other values of $z,z^{\prime}$  also .  However for our purposes the above mentioned one is enough. Note also that here we consider only $z=0$ plasmon contribution believing that it is more important due to the roughness at $z=0$ and not at $z=L$. One can also be convinced that the Green's function Eq.(\ref{magplasmon}) has a pole.  To find the pole, we equate the denominator of Eq.(\ref{magplasmon})to $0$. Getting free from the square roots near the pole values, the Green's function can be represented in the form
\begin{equation}
G_+^0(p|z,0^+)=\frac{a(K_{sp+})e^{i\sqrt{k_+^2-K_{sp+}^2}Z}}{K_{sp+}^2-p^2+i\frac{K_{sp+}}{l_{in+}}},
\label{magpole}
\end{equation}
where
\begin{eqnarray}
K_{sp+}^2=\frac{k_0^2Re\varepsilon_+}{1+Re\varepsilon_+},\quad l_{in+}^{-1}=\frac{k_0Im\varepsilon_+}{Re\varepsilon_+(1+Re\varepsilon_+)},\nonumber \\
a(K_{sp+})=-i\frac{\sqrt{k_+^2-K_{sp+}^2}-\varepsilon_+\sqrt{k_0^2-K_{sp+}^2}}{1-\varepsilon_+^2}.
\label{magpoledet}
\end{eqnarray}
Here $l_{in+}$ describes damping of right-hand magnetoplasmon due to electromagnetic losses and we assume that $Re\varepsilon_+<-1$ and $|Re\varepsilon_+|\gg Im\varepsilon_+$. It follows from Eqs.(\ref{magpole}) and (\ref{cohpart}) that the scattered field and corresponding Faraday angle will resonantly increase provided that one of the inverse lattice wave numbers $K_{nm}$ coincides with the magnetoplasmon wave number $K_{+sp}$, see also \cite{nature}. Using Eqs.(\ref{magpole}),(\ref{magpoledet}) and (\ref{cohpart}) and keeping only the resonance term in the sum of Eq.(\ref{cohpart}), from Eq.(\ref{faranglesc}) one has
\begin{equation}
\tan\theta_p\approx \frac{k_0^2h_0^2(\varepsilon-1)}{2}\left[\frac{l_{in+}a(K_{sp+})}{K_{sp+}}-\frac{l_{in-}a(K_{sp-})}{K_{sp-}}\right],
\label{farangleper}
\end{equation}
where $h_0\equiv h(K_{sp+})$ characterizes the height of periodical grating. To get the plasmon resonance contribution into Faraday rotation angle in the periodical grating case, one has to substitute the parameters $K_{sp\pm}$ and $a(K_{sp\pm})$ from Eq.(\ref{magpoledet}) into Eq.(\ref{farangleper}). The final result, in the limit $g\to 0$, reads as follows
\begin{equation}
\tan\theta_p\approx -\frac{2igk_0h_0\sqrt{\varepsilon}}{Im\varepsilon}.
\label{farperfin}
\end{equation}
Equation (\ref{farperfin}) with Eq. (\ref{ranfin}) (see below) represent the central results of this paper.
The main difference of the plasmon resonance contribution $\tan\theta_p$ compared to smooth surface metallic film contribution, Eq.(\ref{thin}), is that the former dependences on the imaginary part of the dielectric permittivity $Im\varepsilon$. Comparing Eqs.(\ref{thin}) and (\ref{farperfin}), we have
\begin{equation}
\frac{Re\theta_p}{Re\theta_0}\sim\frac{4h_0}{L}\frac{\sqrt{|Re\varepsilon|}}{Im\varepsilon}.
\label{ratioper}
\end{equation}
For nanoscale metallic films, usually $h_0\sim L$. Taking into account that for noble metals in the optical region $|Re\varepsilon|\gg Im\varepsilon$, one has $Re\theta_p\gg Re\theta_0$.

To apply the obtained results to the experiment \cite{nature13}, one can model the composite system consisting of garnet substrate with gold surface profile by an effective metallic film with dielectric permittivity tensor. The diagonal part of the latter is mainly determined by gold (at optical wavelengths its absolute value is much larger than that of garnet) and non-diagonal part determined by bismuth substituted yttrium iron garnet value. Taking at $\lambda=963nm$ \cite{nature13} $Re\varepsilon\approx -40$ , $Im\varepsilon\approx 2.5$ \cite{Christy72} and $h_0\sim L$, from Eq.(\ref{ratioper}) we obtain that the plasmon enhancement factor is of order $10$.
That agrees well with the experimental value \cite{nature13} $8.9$. For the profile periods $a=495nm$ and $b=\infty$  the resonant number is $n=2$. Note that Faraday rotation angle for smooth garnet film, follows from Eqs.(\ref{angleflat},\ref{thin}). Taking $g=0.016$, $\varepsilon=6.7$, $\lambda=963nm$, $L=150nm$ \cite{nature13}, one has $\theta_0\sim 0.16deg$.

For the large $n,m$ one can not separate out a single resonance term from the sum over the inverse lattice wave numbers, Eq.(\ref{cohpart}). In this case the summation can be replaced by integration ($\sum_{K}\to \frac{S}{(2\pi)^2}\int d\vec K$, where $S$ is the area of the system). Carrying out the integration over $\vec K$ and making use of Eqs.(\ref{backsol},\ref{reftrans},\ref{faranglesc}), one finds
for the real part of Faraday angle, in the limit $L\to 0$, the following expression ($G_{\pm}(\vec\rho,\vec\rho)\equiv G_{\pm}(\vec\rho,\vec \rho,0^+,0^+)$ )
\begin{equation}
Re\tan\theta_p=\frac{k_0^2(\varepsilon-1)h_0\bigg[ImG_+(\vec\rho,\vec\rho)-ImG_-(\vec\rho,\vec\rho)\bigg]}{2}.
\label{percont}
\end{equation}
This is a general expression, independent of the surface periodic profile model.
For simplicity we discuss only constant harmonic grating case, i.e., $h(\vec K)=const=h_0$.
The quantities $Im G_{\pm}(\vec\rho, \vec\rho)$ are the local density of states of right hand and left hand polarized magnetoplasmons. Because of translational invariance, they depend only on the difference of the arguments and therefore are independent of local point $\rho$. Expanding $Im G_{\pm}(\vec\rho, \vec\rho)$ on $g$, one gets that $Re\tan\theta_p\sim g\partial ImG/\partial\varepsilon$. Thus, the measurement of the  rotation angle  gives information on the density of states  \cite{AGbrus}. More as a consequence of the periodicity of $h(\vec\rho)$, the plasmon spectrum consists of energetic bands and gaps. The above mentioned derivative gets its maximal values at the edges of these bands. Similar behavior for the Faraday rotation was found in 1D periodical systems \cite{KHB08,GGB13}.

\subsection{ Random surface profile}
Now consider the case when the surface profile is random. We assume that $h(\vec\rho)$ is a Gaussian distributed random function
 \begin{equation} <h(\vec\rho)h(\vec\rho^{\prime})>=h^2\sigma^2\delta(\vec\rho-\vec\rho^{\prime}),\quad <h(\vec\rho)>=0
 \label{ranprofile}
 \end{equation}
 where $<...>$ denotes the ensemble average and $h$ and $\sigma$ are the root-mean-square roughness and correlation length, respectively. In order to average the Faraday angle over the realizations of random roughness and to separate its real part, it is convenient to multiply the numerator and denominator of Eq.(\ref{detangle}) by $H_+^*(L)+H_-^*(L)$, see also \cite{GG13}
 \begin{equation} <\tan\theta>\approx-i\frac{<(H_{+}(L)-H_{-}(L))(H_+^*(L)+H_-^*(L))>}{<(H_{+}(L)+H_{-}(L))(H_+^*(L)+H_-^*(L))>}.
 \label{multiply}
 \end{equation}
Like in the periodical grating case, we decompose the magnetic field into background and scattered parts $H_{\pm}=H_{\pm}^0+H_{\pm}^s$. In the denominator of Eq.(\ref{multiply}) for small roughness $h\to 0$, one can keep only the terms containing background fields $H_{\pm}^0$. In respect to the numerator, one should keep only the terms containing the scattered fields because the terms associated with the background field are small for thin films: $H_+^0(L)-H_-^0(L)\sim L\to 0$. For the real part of plasmon diffusional contribution to the Faraday angle, one finds from Eq.(\ref{multiply})
\begin{equation} Re<\tan\theta>^D=-i\frac{<H_+^sH_-^{*s}>-<H_-^sH_+^{*s}>}{|H_+^0|^2+|H_0^0|^2+H_+^0H_-^{*0}+H_-^0H_+^{*0}}.
\label{realdif}
\end{equation}
Recall that the scattered field is determined by Eq.(\ref{scafield}) and that the terms $<|H_{\pm}^s|^2>$ do not contribute to the real part of Faraday angle because the denominator of Eq.(\ref{realdif}) is real. To find the averages in Eq.(\ref{realdif}) one needs the averaged over random roughness Green's function, taking into account the plasmon multiple scattering effects on the surface roughness (see Refs. \cite{AZ83,ASB85}). Random roughness leads to damping of surface magnetoplasmon due to elastic scattering. The final answer for the magnetoplasmon Green's function, averaged over the randomness, reads
\begin{equation}
G_+(p|0^+,0^+)=\frac{a_+}{K_{sp+}^2-p^2+iK_{sp+}/l_+},
 \label{avgrfun}
\end{equation}
where magnetoplasmon elastic mean free path is determined as
\begin{equation}
 l_+=\frac{4K_{sp+}}{\beta a_+^2}
 \label{elmean}
 \end{equation}
and $\beta=k_0^4(\varepsilon-1)^2h^2\sigma^2$. In Eq.(\ref{avgrfun}) we neglect $l_{in}^{-1}$ compared to $l_+^{-1}$. We will take the contribution of the former into account in the diffusional propagator (see below). In the weak scattering limit $K_{\pm sp}l_{\pm}\gg 1$, the main contribution to the average quantities in Eq.(\ref{realdif}) gives the magnetoplasmons' diffusion. Using Eq.(\ref{scafield}) one can represent the diffusional contribution in the form
\begin{eqnarray}
&<H_+^s(L)H_-^{*s}(L)>^D=\beta H_+^0(0)H_-^{*0}(0)P_{+-}(K=0)\nonumber \\
&\int\frac{d\vec p}{(2\pi)^2}G_+(\vec p)G_-^*(-\vec p)\int\frac{d\vec p}{(2\pi)^2}G_+(\vec p|L,0)G_-^*(-\vec p|0,L)\nonumber, \\
 \label{difcontr}
\end{eqnarray}
where $G_+(\vec p)\equiv G_+(\vec p|0^+,0^+)$, $P_{+-}$ magnetoplasmon diffusion propagator  which is determined by ladder diagrams presented in Fig.2 (see for example \cite{Mackjohn88,Theo99})
 \begin{figure}
\vspace{-2cm}
 \begin{center}
\includegraphics[width=10cm]{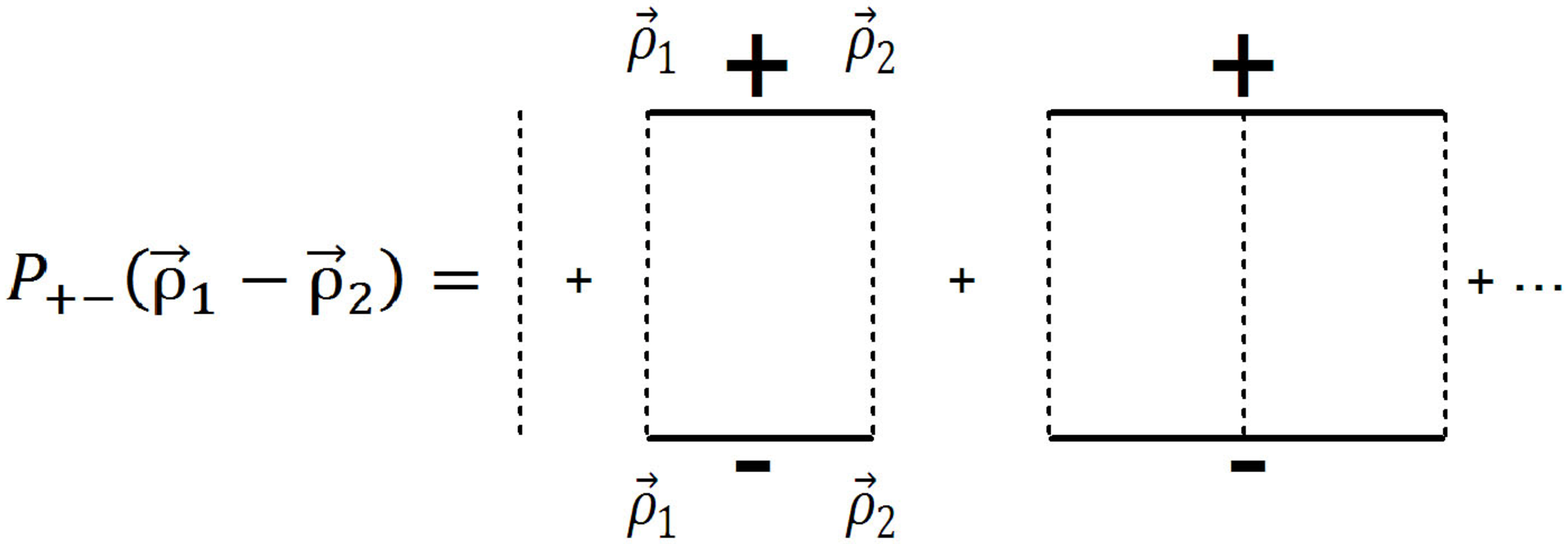}
\caption{Ladder diagrams. Upper line is the right hand polarized magnetoplasmon averaged Green's function, bottom line is the left hand polarized magnetoplasmon Green's function and the dashed line is the random field correlation function $\beta\delta(\vec\rho_1-\vec\rho_2).$}
\label{fig.2}
\end{center}
\end{figure}
Summing the ladder diagrams in the limit $g\to 0$, one has
\begin{equation}
P_{+-}(K)=\frac{8\beta}{\frac{l}{l_{in}}+K^2l^2-i\frac{gk_0l}{2(\varepsilon+1)^2}\sqrt{\frac{\varepsilon+1}{\varepsilon}}}.
\label{difprop}
\end{equation}
Expression (\ref{difprop}) was derived in the limits $Kl\ll 1$ and $\lambda\ll l\ll l_{in}$. The propagator $P_{-+}(K)$ is obtained from $P_{+-}(K)$ by changing the sign of $g$. Calculating the integrals in Eq.(\ref{difcontr}) in the limits $g\to 0$ and $L\to 0, k_0L\sqrt{|\varepsilon|}\ll 1$ , we arrive at
\begin{equation}
<H_+^s(L)H_-^{*s}(L)>^D=\frac{P_{+-}(K=0)}{\beta}.
\label{difav}
\end{equation}
Finally, using Eqs.(\ref{realdif}), (\ref{difprop}) and (\ref{difav}), for diffusional contribution to the Faraday angle, we obtain
\begin{equation}
<Re\tan\theta>^D=\frac{2gK_{sp}l_{in}^2}{\varepsilon(\varepsilon+1)l}.
\label{ranfin}
\end{equation}
Note, that the ratio $l_{in}/l$ has is the average number of scatterings of plasmon. A similar result for Faraday rotation in 3D disordered medium is obtained in \cite{GG13}. However in 2D systems, Faraday rotation is more sensitive to the number of scatterings (square dependence against the linear in 3D case) as well as to dielectric permittivity of thin film. Near the surface plasmon resonance $\varepsilon(\omega)+1=0$, the Faraday angle enhances because $<Re\tan\theta>^D\sim 1/\sqrt{|\varepsilon+1|}$. Comparing Eq. (\ref{thin}) with the flat surface contribution, Eq.(\ref{angleflat}), we have
\begin{equation}
\frac{<Re\tan\theta>^D}{Re\tan\theta_0}\sim\frac{1}{\sqrt{Re\varepsilon(Re\varepsilon+1)^3}}\frac{l_{in}^2}{Ll}.
\label{lastexpr}
\end{equation}
If the diffusion of magnetoplasmon is realized on the film surface, i.e., the inequality $\lambda\ll l\ll l_{in}$ is met , then the condition $l_{in}^2\gg Ll$ should hold. As a consequence, the diffusion contribution to the Faraday rotation angle can be the dominant one. Now let us make some numerical estimates to clarify whether or not the mentioned inequality takes place.

Assuming that the roughness is created by the randomly adsorbed on the surface nanoparticles, we have: $h\sim \sigma\sim r$, where $r$ is the radius of a nanoparticle. For gold nanoparticles with radius $r=30nm$ at $\lambda=600nm$ \cite{Tkachuk11}, $Re\varepsilon=-7.8$, $Im\varepsilon=1.6$ \cite{Christy72} and $\beta=k_0^4(\varepsilon-1)^2h^2\sigma^2\approx 0.75$. The surface plasmon wave number $K_{sp}$ and the constant $a$ in Eq.(\ref{magpoledet}) are estimated as $k_0$ and $0.1k_0$, respectively. The surface plasmon elastic mean free path $l$ is found from Eq.(\ref{elmean}) to be approximately $l\sim 97\lambda$. If the losses are caused by the gold substrate, then the inelastic mean free path can be estimated using Eq.(\ref{magpoledet}) and the above mentioned numbers, $l_{in}\sim 5.3\lambda$ . So plasmon diffusion inequalities $\lambda\ll l\ll l_{in}$ are not realized in ordinary conditions.

However, close to the nanoparticle surface plasmon resonance the physical situation is completely different, because the expression Eq.(\ref{elmean}), for the determination of the plasmon elastic mean free path is not valid any more. In this case the elastic mean free path can be estimated as $l=L/n_s\sigma$, where $n_s$ is the surface concentration of nanoparticles and $\sigma$ is the cross section of the interaction of the surface plasmon with the nanoparticle. When the surface plasmon wave number coincides with the nanoparticle surface plasmon resonance wave number, the plasmon elastic cross section resonantly enhances up to several orders compared to ordinary situation, see for example  \cite{luk07}. Therefore, close to the resonance, the magnetoplasmon elastic mean free path becomes essentially smaller and the condition of its diffusion can be easily fulfilled.

\section {Summary} We have considered the Faraday rotation of light passing through a thin metallic film with nanostructured surface grating. In the periodical grating case, the enhancement of the Faraday angle happens when the surface plasmon wave number coincides with one of the wave vectors of inverse lattice, characterizing the grating periods on the surface.  In the random surface profile case, the dominant contribution to the Faraday angle gives the diffusion of  magnetoplasmons. If the random roughness is created by the randomly embedded nanoparticles on the surface, then the maximum Faraday rotation angle of the transmitted wave is achieved when surface plasmon wave number coincides with the nanoparticle surface plasmon resonance wave number. Experimental manifestations of the obtained results are discussed.

{\bf Acknowledgments.}

We are grateful to O. del Barco for preparing the figures. We thank W. Whitaker for a critical reading of the mansucript. V.G. acknowledges
partial support by FEDER and the Spanish DGI under Project No. FIS2010-16430.


\begin{thebibliography}{99}
\bibitem{uchida09} H. Uchida, Y. Masuda, R.Fujikawa, A.V.Baryshev, M.Inoue, Journal of Magnetism and Magnetic Materials {\bf 321} 843 (2009).
\bibitem{Jain09}  Prashant K.Jain, Yanhong Xiao, Ronald Walswort, and Adam E.Cohen, Nano Lett.,{\bf 9(4)} 1644 (2009).
\bibitem{Tkachuk11} S.Tkachuk, G.Lang, C.Krafft and I.Mayergoyz, Journal of Applied Physics, {\bf 109} 07B717 (2011).
\bibitem{Chikan11} Raj Kumar Dani, Hongwang Wang, Stefan H.Bossmann, Gary Wysin and Viktor Chikan, The Journal of Chemical Physics, {\bf 135} 224502 (2011).
\bibitem{natphys11} I. Crassee, J. Levallois, A. L. Walter, M. Ostler, A. Bostwick, E. Rotenberg, T. Seyller, D. van der Marel, and A. B.Kuzmenko,
Nat. Phys. 7, 48 (2011).
\bibitem{Chikan13} G. M. Wysin, Viktor Chikan, Nathan Young, and Raj Kumar Dani, J. Phys.: Condens. Matter {\bf 25} 325302 (2013).
\bibitem{nature13} Jessie Yao Chin, Tobias Steinle, Thomas Wehlus, Daniel Dregely, Thomas Weiss, Vladimir I. Belotelov, Bernd Stritzker and Harald Giessen, Nature Communications,{\bf DOI: 10.1038/ncomms2609} (2013) 1-6.
\bibitem{Hay08} K.Hayashi, R.Fujikawa, W.Sakamoto, M.Inoue and T.Yogo, J.Phys.Chem.C{\bf 112} 14255 (2008).
\bibitem{Yu08} H.C.Y.Yu, M.A.Eijkelenborg, S.G.Leon-Saval, A.Argyros and G.W.Barton, Appl.Opt. {\bf 47}, 6497 (2008).
\bibitem{Liu12} F.Liu, T.Makino, T.Yamasaki, K.Ueno, A.Tsukazaki, T.Fukumura, Y.Kong and M.Kawasaki, Phys.Rev.Lett. {\bf 108}, 257401 (2012).
\bibitem{Raether88} H.Raiether, {\it Surface Plasmons on Smooth and Rough Surfaces and on Gratings} (Springer Tracts in Modern Physics vol.111) 1988 Berlin,Springer.
\bibitem{GG13} V. Gasparian and Zh. S. Gevorkian, Phys.Rev. A {\bf 87}, 053807(2013).
\bibitem{arkady} A.G. Aronov, V.M. Gasparian, and Ute Gummich, J. Phys. Condens. Matter; {\bf 3}; 3023 (1991).
\bibitem{LL} L.D.Landau, and E.M.Lifshitz, {\it Electrodynamics of Continuous Media} (Pergamon Press, 1982).
\bibitem{nature} William L. Barnes, Alain Dereux and Thomas W. Ebbesen, Nature {\bf 424} 824 (2003).
\bibitem{AZ83} K.Arya and R.Zeyher, Phys.Rev.B {\bf 28}, 4090,(1983).
\bibitem{ASB85} K.Arya, Z.B.Su and Joseph L.Birman, Phys.Rev.Lett.{\bf 54}, 1559,(1985).
\bibitem{Christy72} P.B.Johnson and R.W.Christy, Phys. Rev. B {\bf 6}, 4370,(1975).
\bibitem{AGbrus} A.G. Aronov, and V.M. Gasparian, Solid St. Commun. {\bf 73}, 61 (1990).
\bibitem{KHB08} A.B.Khanikaev, A.B.Baryshev, P.B.Lim, H.Uchida, M.Inoue,A.G.Zhdanov,A.A.Fedyanin,A.I.Maydykovskiy and O.A.Aktsipetrov, Phys.Rev. B {\bf 78},193102,(2008).
\bibitem{GGB13} V. Gasparian, Zh. Gevorkian and O. del Barco, Phys. Rev. A {\bf 88} 023842 (2013).
\bibitem{Mackjohn88} F.C.MacKintosh and S.John, Phys. Rev. B {\bf 37} 1884 (1988).
\bibitem{Theo99} M.C.W.van Rossum and Th.M.Nieuwenhuizen, Rev. Mod. Phys. {\bf 71} 313 (1999).
\bibitem{luk07} B.S.Luk'yanchuk, M.I.Tribelsky, V.Ternovsky, Z.B.Wang, M.N.H.Hong, L.P.Shi and T.C.Chong, J.Opt.A:Pure Appl.Opt.{\bf 9} 294 (2007).
\end{thebibliography}
\end{document}